# Hybrid Metaheuristic Optimization of Distributed Control System Hardware Architecture with Model-Based Verification

**Ruslan Zakirzyanov***

** NEXT engineering, LLC*
*Kazan, Russia, (zr@nexteng.ru)*

**Abstract**: Large-scale chemical plants rely on distributed process control systems (PCS) comprising numerous processing units, communication modules, and I/O devices interconnected via industrial networks. The design of a cost-efficient and reliable hardware architecture under partial uncertainty in plant parameters remains a challenging combinatorial optimization problem. This paper proposes a formal model for distributed control system hardware architecture synthesis. A hybrid ant colony-based metaheuristic framework is developed to construct feasible hierarchical architectures. The proposed approach is validated on a large-scale sulfuric acid plant control system case study. Plant parameters are identified from operational data, system stability is analyzed, and a controller synthesis is performed based on the optimized architecture. The results demonstrate the feasibility of the approach and confirm that the obtained architecture satisfies structural and dynamic performance requirements.

## 1. INTRODUCTION

Modern PCS deployed at large industrial chemical plants are constructed using commercially available components, including controllers, input/output modules, and network switches. Since the characteristics of these devices are predefined, overall system performance is largely determined by the hardware architecture. Such architectures are typically hierarchical, comprising several levels from field devices at the bottom to human–machine interface (HMI) systems at the supervisory level. In practice, PCS architectures are often designed based on engineering experience or manufacturer recommendations and are therefore not guaranteed to be cost-optimal.

This paper addresses the problem of constructing an optimal hierarchical PCS hardware architecture under multiple structural and resource constraints. The objective function is the total system cost, defined as the sum of the costs of all hardware components. Signal allocation, device capacity limits, and hierarchical connectivity constraints must be satisfied simultaneously. The resulting optimization problem is combinatorial and NP-hard.

Structural optimization of technical systems has been widely studied. Bussemaker et al. (2025) investigate structural optimization problems for aerospace systems. Sinha et al. (2025) consider structural optimization problems for railway systems using a design structure matrix (DSM) framework. Both exact and approximate optimization methods have been proposed for related classes of structural design problems.

In recent decades, metaheuristic algorithms have been extensively used for solving high-dimensional constrained optimization problems. Widely applied methods include Simulated Annealing (Kirkpatrick et al., 1983), Particle Swarm Optimization (Kennedy et al., 1995), Grey Wolf Optimizer (Mirjalili et al., 2014), Tabu Search (Glover, 1986), and Ant Colony Optimization (Dorigo et al., 2004). Metaheuristics are also employed in industrial engineering, for example for PID controller parameter tuning (Olmez et al. 2025) and controller placement (Frdiesa, 2024). Most existing studies address general structural optimization problems, which are difficult to directly apply to PCS hardware architecture design. Zakirzyanov (2025a) proposed a model for constructing an adaptive structure of the PCS and investigated the possibilities of applying metaheuristics to it. However, the proposed model did not adequately reflect real-world device constraints, and dynamic verification was not performed.

This paper proposes an improved model and a hybrid algorithm combining deterministic construction procedures with a metaheuristic search mechanism. The deterministic component ensures hierarchical feasibility of the architecture, while the metaheuristic component is responsible for device selection within the hierarchy. This approach reduces the number of infeasible candidate solutions compared to direct metaheuristic application. To improve algorithm robustness, adaptive parameter tuning and a local search procedure are incorporated to mitigate stagnation in local minima.

Extensive computational experiments were conducted to evaluate the proposed method. The algorithm was applied in the design of a distributed control system. The optimized architecture was incorporated into the dynamic mathematical model of the production section. Model parameters were identified using real operational data. Closed-loop simulations confirmed that the resulting system satisfies prescribed control performance requirements.

The main contributions of this paper are:

(i) an improved formal optimization model for hierarchical PCS hardware architecture synthesis under structural and resource constraints;

(ii) a modified hybrid deterministic-metaheuristic algorithm for feasible architecture construction;

(iii) parameter tuning and adaptive heuristics aimed at improving solution feasibility and search efficiency;

(iv) experimental validation on a real sulfuric acid production control system including dynamic model identification and closed-loop verification.

The remainder of the paper is organized as follows. Section 2 formulates the optimization problem. Section 3 presents the proposed hybrid algorithm. Section 4 reports computational and industrial case study results. Section 5 addresses dynamic verification and controller synthesis. Section 6 concludes the paper.

## 2. PROBLEM FORMULATION

Zakirzyanov (2025b) proposed a detailed formal description of the industrial PCS architecture optimization problem. The model is extended to overcome several limitations of the previous formulation, including a fixed processor level assumption, representation in terms of control loops instead of physical signals, and restriction to a single interface type. The main new model statements are presented in this section.

The PCS architecture is modeled as a rooted directed tree $G = (\mathrm{V}, \mathrm{E})$, where $V$ is the set of nodes representing devices $v \in V$ and $E$ is the set of edges representing communication links between devices. An example of such a tree is shown in Fig. 1. The number of levels in the architecture is specified by the designer.

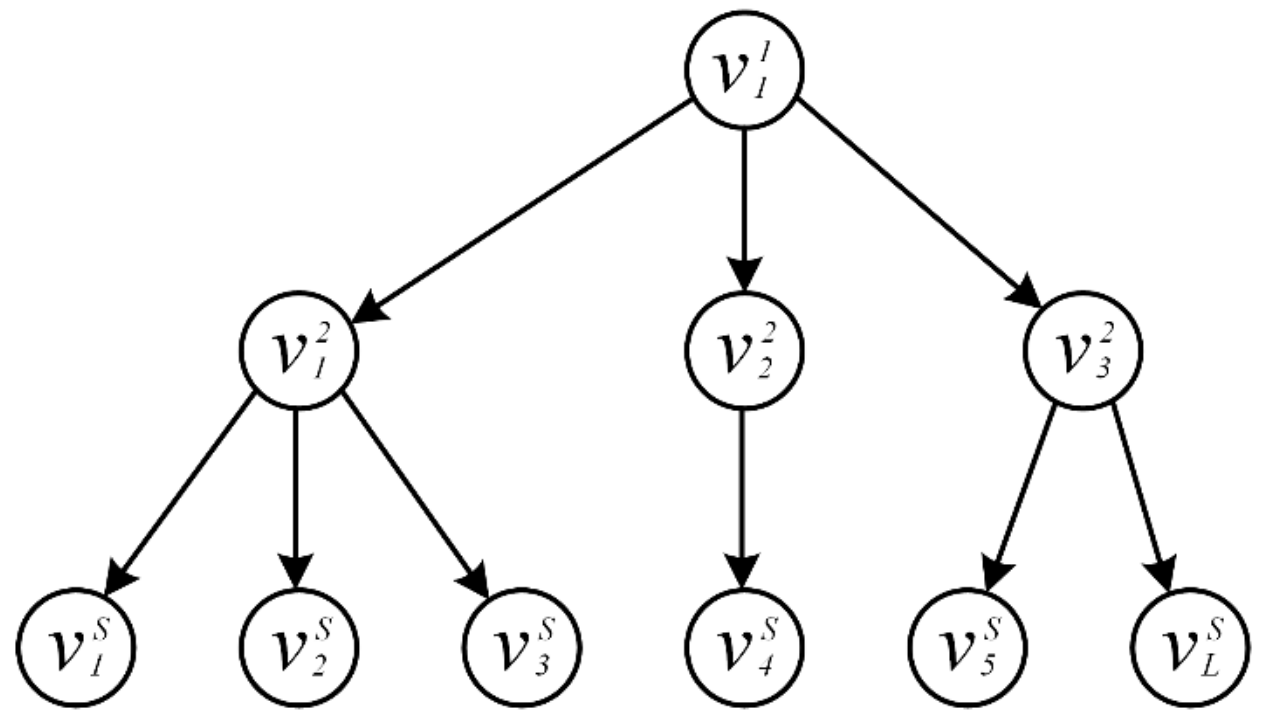

Figure 1. Hierarchical architecture of the PCS.

Let $\mathcal{B} = \{1, \dots, B\}$ denote signal types, where $B \in \mathbb{N}$, and $\mathcal{F} = \{1, \dots, F\}$ denote interface types, where $F \in \mathbb{N}$. Each signal type $j \in \mathcal{B}$ is characterized by the pair:

$$b_j = (r_j, w_j), \quad \mathrm{j} = 1, \dots, B, \tag{1}$$

where $r_j \in \mathbb{R}_+$ denote the signal variable memory, and $w_j \in \mathbb{R}^+$ denote the amount of program instructions for the signal processing. Each signal requires one appropriate device channel for connection.

All the signals are grouped into the control rooms. Let $\mathcal{H} = \{1, \dots, H\}$ denote control rooms, where $H \in \mathbb{N}$. The set of the physical signals is defined as $\mathcal{A} = \{a_1, \dots, a_{|\mathcal{A}|}\}$. Each signal $a_k \in \mathcal{A}$ is defined as:

$$a_k = (a_k^1, a_k^2),\ a_k^1 \in \mathcal{B},\ a_k^2 \in \mathcal{H}\ \ k = 1, \dots, |\mathcal{A}|, \tag{2}$$

Let $\mathcal{U} = \{u_1, \dots, u_{|\mathcal{U}|}\}$ denote the set of available device types. Devices are divided into two roles: processors and repeaters. Each type $u_i \in \mathcal{U}$ of structure node is characterized by the parameter vector:

$$u_i = (C_i, R_i, P_i, T_i, y_i, \tau_i, \gamma_i, \boldsymbol{M}_i, \boldsymbol{N}_i), \quad i = 1, \dots, |\mathcal{U}|, \tag{3}$$

where $C_i \in \mathbb{R}^+$ – device cost, $R_i \in \mathbb{R}^+$ – memory capacity, $P_i \in [0,1]$ – failure probability, $T_i \in \mathbb{R}^+$ – instruction execution time, $y_i \in \{0,1\}$ – device role (0 – repeater, 1 – processor), $\tau_i \in \mathbb{R}^+$ – transmission delay for repeaters, $\gamma_i \in \mathcal{F}$ – upper connection interface type, $\boldsymbol{M}_i = (M_{i1}, \dots, M_{iF})$ – maximum number of child devices for each interface type, $\boldsymbol{N}_i = (N_{i1}, \dots, N_{iB})$ – maximum number of physical channels for each channel type. Multiple instances of each device type may be selected.

$S \in \mathbb{N}$ is the number of tree levels (1 is the tree root, $S$ is the leaves level). Field signals can be connected only to the leaf level devices.

Binary decision variables are introduced to describe signal allocation: $x_{va}, z_{va}, g_{va} \in \{0,1\}$. Variable $x_{va} = 1$ if signal $a$ is physically connected to leaf node $v$, and 0 otherwise. Variable $z_{va} = 1$ if signal $a$ is processed at node $v$, and 0 otherwise. Variable $g_{vh} = 1$ if device $v$ belongs to the control room $h$, and 0 otherwise.

The model includes constraints ensuring signal allocation consistency, device memory capacity limits, execution time constraints, reliability requirements, hierarchical connectivity feasibility.

The objective is to determine an optimal hierarchical architecture $G^*$ that minimizes the total system cost:

$$C^* = \min_G \textstyle\sum_{v \in V} C_v, \tag{4}$$

subject to the imposed constraints.

## 3. PROPOSED METHOD

The PCS architecture synthesis problem is a constrained combinatorial optimization task. Direct application of population-based metaheuristics to this problem often generates a large fraction of infeasible solutions, resulting in excessive computational cost due to repeated constraint violations. To address this issue, we propose a hybrid framework that combines a deterministic tree-building procedure with a metaheuristic device selection mechanism. A deterministic construction algorithm incrementally builds a tree-shaped architecture. The procedure starts by constructing the main trunk (root-to-leaf path) of the tree. After the trunk is formed, physical signals are allocated to leaf-level devices. If the current architecture cannot accommodate all signals, additional devices and branches are added where needed.

Because the construction explicitly enforces structural constraints at each step, infeasible architectures are avoided by design. Device selections are performed using ant colony optimization (ACO).

Dorigo et al. (2004) presented a detailed description of the ACO algorithm. ACO is a graph-oriented metaheuristic and is well suited for constructive decisions in hierarchical architectures. At each decision point, the next device type is selected probabilistically from the set of admissible candidates $\mathcal{U}$ according to:

$$P_i = \frac{\tau_i^{\alpha}\eta_i^{\beta}}{\sum_{k\in\mathcal{U}_{adm}}\tau_k^{\alpha}\eta_k^{\beta}}, \tag{5}$$

where $P_i$ is the probability of choosing candidate $i$, $\tau_i$ is the pheromone value associated with $i$, $\eta_i$ is the heuristic desirability, $\alpha$ is the pheromone weight, $\beta$ is the heuristic weight, $\mathcal{U}_{adm} \subseteq \mathcal{U}$ denotes admissible candidate device types.

After each iteration, pheromone values are updated using evaporation and deposition:

$$\tau_i \leftarrow (1-\rho)\tau_i + \Delta\tau_i, \tag{6}$$

where $\rho \in (0,1)$ is the evaporation rate and $\Delta\tau_i$ is proportional to the inverse cost of the constructed architecture.

A commonly used baseline heuristic favors low-cost devices:

$$\eta_i = \frac{1}{C_i}, \tag{7}$$

where $C_i$ is the cost of candidate device type *i*. However, cost-only heuristics may lead to structural infeasibility states near feasibility boundaries, for example, by selecting devices that are inexpensive but provide insufficient capacity for further expansion.

To improve feasibility, we propose a level-dependent adaptive heuristic that accounts for the role of a device within the hierarchy. Let s denote the current level (s $\in \{1,\dots,S\}$). The heuristic prioritizes channel capacity to better accommodate remaining signals for leaf level:

$$\eta_i = \frac{N_i^2}{C_i}, \quad s = S, \tag{8}$$

where $N_i = \sum_{b=1}^{B} N_{ib}$ denotes total channel capacity.

For internal levels ($s < S$), the heuristic additionally prioritizes branching capability via the maximum number of child devices $M_i$:

$$\eta_i = \frac{N_i^2 M_i}{C_i}, \quad s < S, \tag{9}$$

where $M_i = \sum_{f=1}^{F} M_{if}$ denotes total child devices capacity.

Furthermore, the heuristic is adaptively adjusted using the number of unassigned signals $A_{left}$ and the available structural expansion capacity (free child slots) $B_{free}$. When $A_{left} > B_{free}$, candidates with insufficient branching capability are penalized to prevent premature saturation of the hierarchy.

The performance of ACO is sensitive to its parameters, particularly $\alpha$, $\beta$, and $\rho$. These parameters affect both feasibility (the share of successful constructions) and stability (solution variability). Therefore, parameter tuning is formulated as a bi-objective optimization problem maximizing feasibility rate and minimizing solution variance. This problem can be solved using Pareto-based tuning, allowing the designer to trade off feasibility and stability. We performed bi-objective parameter tuning using NSGA-II (Deb et al., 2002). Following parameter tuning, a local search (LS) improvement stage is applied to the best constructed architecture using a simple device-replacement neighborhood: a randomly selected node is replaced with an alternative device type of the same hierarchical level, and the change is accepted if it improves the objective while preserving feasibility.

## 4. EXPERIMENTAL VALIDATION

A dedicated Python implementation was developed for the computational experiments. All experiments were performed on a workstation (Intel Core i5/16GB/Windows 10). A real-world PCS design project for a large-scale sulfuric acid plant was used as an industrial case study. The control system consisted of 1,068 physical signals of four types (AI, DI, AO, DO). Three interface types, listed in Table 1, were adopted for connecting the devices.

Table 1. Interface types

| No | Name/Protocol | Cycle time, ms |
|---|---|---|
| 1 | RS-485/Modbus RTU | 35 |
| 2 | Ethernet/Modbus TCP | 10 |
| 3 | EtherCAT | 2 |

Various commercially available components were considered for the system's construction. Some of these are listed in Table 2. The table includes the central processor modules, input/output modules, and network switches.

Table 2. Device types

| No | Name | $N_i$ | $M_{i1}+M_{i2}+M_{i3}$ | $\gamma_i$ |
|---|---|---|---|---|
| 1 | CU 00 062 | - | 4+0+40 | 2 |
| 2 | CU 00 021 | - | 2+0+10 | 2 |
| 3 | AI 08 031 | 8 | - | 3 |
| 4 | AI 16 012 | 16 | - | 3 |
| 5 | DI 16 032 | 16 | - | 3 |
| 6 | DI 32 013 | 32 | - | 3 |
| 7 | DO 16 021 | 16 | - | 3 |
| 8 | DO 32 031 | 32 | - | 3 |
| 9 | AO 08 031 | 8 | - | 3 |
| 10 | 8 port Switch | - | 0+8+0 | 2 |
| 11 | 5 port Switch | - | 0+4+0 | 2 |
| 12 | AI8 | 8 | - | 1 |
| 13 | DI8 | 8 | - | 1 |
| 14 | DO4 | 4 | - | 1 |
| 15 | AO2 | 2 | - | 1 |

The best-found architecture contains 48 devices. The algorithm convergence plot is shown in Fig. 2. The optimized values of the algorithm parameters, selected from the Pareto front (maximizing feasibility rate and minimizing solution variance), are equal to $\alpha = 2.45, \beta = 3.97, \rho = 0.79$. The resulting structure has four hierarchical levels.

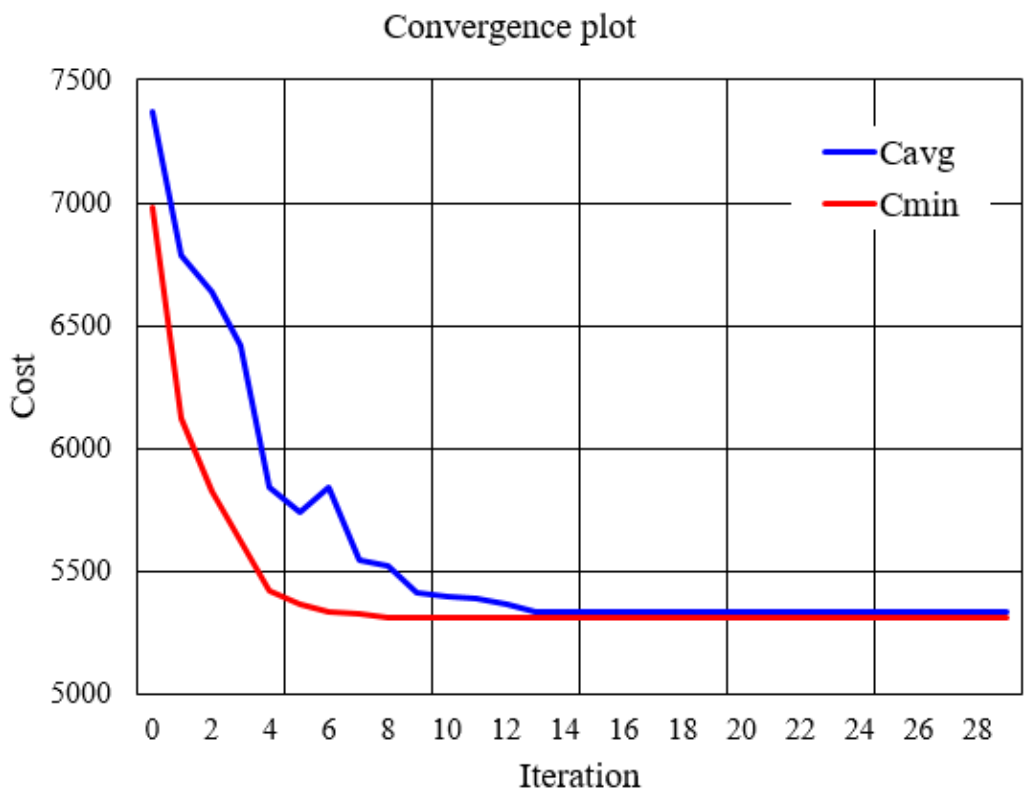

Figure 2. Convergence plot.

In Fig. 2, the red line shows the best-so-far minimum over 20 runs, while the blue line shows the average best-so-far value. Obtained average runtime $T = 17.2$s, feasibility rate $W = 50.11\%$.

## 5. DYNAMIC ANALYSIS

The resulting four-level architecture was incorporated into the dynamic model of the networked control system. Communication channels and devices introduce delays consisting of a fixed transport delay and a stochastic component associated with packet losses. A subsection of the sulfuric acid circulation process, comprising a drying tower, a collector, and a heat exchanger, was selected for validation. The controlled variable is the sulfuric acid temperature, regulated by two control valves. A sixth-order state-space model was constructed for the control plant. The model parameters were identified using the subspace state-space identification method (N4SID) based on archived industrial process data. The initial realization was refined using prediction error minimization (PEM). Measured actuator signals were used as model inputs to avoid closed-loop identification bias. Closed-loop simulations indicate that the system satisfies control performance requirements, with overshoot below 16% and settling time = 4.3s. Closed-loop stability was analyzed using vector Lyapunov functions, and an LQR controller was synthesized.

## 6. CONCLUSIONS

The proposed hybrid algorithm enables the synthesis of cost-efficient hierarchical PCS architectures using commercially available components under partial uncertainty in plant dynamics and communication characteristics. The deterministic construction phase ensures structural feasibility, while the adaptive ACO-based search improves solution quality and robustness. The approach was validated on a real sulfuric acid production system, and dynamic verification confirmed satisfactory closed-loop performance of the optimized architecture.

Future work will focus on further refinement of the optimization model, formal scalability analysis, and comparative evaluation against exact and alternative metaheuristic methods on additional industrial case studies.